%% file: main.tex
\documentclass[fleqn,10pt]{wlscirep}
\usepackage[utf8]{inputenc}
\usepackage[T1]{fontenc}
\usepackage{amssymb}
\usepackage{pifont}

\usepackage{soul}
\usepackage{xcolor}
\usepackage{tabularx}
\usepackage{float}
\usepackage[toc,page]{appendix}
\usepackage{algpseudocode}
\usepackage{algorithm}
\usepackage{arydshln}

\newcommand{\toadd}[1]{\textcolor{red}{#1}}

\begin{document}

\title{``Can't Take the Pressure?'': Examining the Challenges of Blood Pressure Estimation via Pulse Wave Analysis}

\author[1]{Suril Mehta}
\author[2]{Nipun Kwatra}
\author[3]{Mohit Jain}
\author[4]{Daniel McDuff}
\affil[1]{Microsoft Research}
\affil[2]{Microsoft Research}
\affil[3]{Microsoft Research}
\affil[4]{Microsoft Research}


\begin{abstract}
The use of observed wearable sensor data (e.g., photoplethysmograms [PPG]) to infer health measures (e.g., glucose level or blood pressure) is a very active area of research. Such technology can have a significant impact on health screening, chronic disease management and remote monitoring. A common approach is to collect sensor data and corresponding labels from a clinical grade device (e.g., blood pressure cuff), and train deep learning models to map one to the other. Although well intentioned, this approach often ignores a principled analysis of whether the input sensor data has enough information to predict the desired metric.
We analyze the task of predicting blood pressure from PPG pulse wave analysis. Our review of the prior work reveals that many papers fall prey data leakage, and unrealistic constraints on the task and the preprocessing steps. We propose a set of tools to help determine if the input signal in question (e.g., PPG) is indeed a good predictor of the desired label (e.g., blood pressure). Using our proposed tools, we have found that blood pressure prediction using PPG has a high multi-valued mapping factor of 33.2\% and low mutual information of 9.8\%. In comparison, heart rate prediction using PPG, a well-established task, has a very low multi-valued mapping factor of 0.75\% and high mutual information of 87.7\%. We argue that these results provide a more realistic representation of the current progress towards to goal of wearable blood pressure measurement via PPG pulse wave analysis.
\end{abstract}

\flushbottom
\maketitle
\thispagestyle{empty}

\input{content/1_introduction}
\input{content/3_motivation}
\input{content/7_discussion}

\section*{Author Contributions}
SM performed analyses, designed experiments, and wrote the manuscript.
NK designed experiments and wrote the manuscript. MJ designed experiments and wrote the manuscript. DM designed experiments and wrote the manuscript.

\section*{Competing Interests}
The Authors declare no Competing Financial or Non-Financial Interests.

\begin{appendices}
\input{content/X_Appendix}

\end{appendices}

\clearpage
\bibliography{FinalSubmission/references}
\end{document}

%% file: content/1_introduction.tex
\section{Introduction}
\footnote{For code see our project page: \url{https://github.com/lirus7/PPG-BP-Analysis}}
The COVID-19 pandemic has highlighted the acute need for technology to support remote health care~\cite{bhat-telehealth-cscw, covid-19-2}. Consultancy McKinsey~\cite{article_health_4} reported a 40-fold increase in the use of telehealth services and a 40\% increase in consumer interest in virtual health solutions when compared to pre-COVID-19 statistics. To provide an example, the ability to estimate vital signs from sensors available in smartphones and wearable devices could have a significant impact on effective management of diseases (e.g., COVID-19, hypertension, diabetes). Frequent measurement of physiological parameters can help in managing medication dosages and understanding the effects of lifestyle changes on health. 

The estimation of vital signs traditionally relies on customized sensors that measure physical or chemical properties of the body. For example, digital sphygmomanometers use sensors to measure the oscillations in the arteries to quantify blood pressure. Although accurate, such medical devices are far from ubiquitous, often are not easy to access and are uncomfortable to use for extended periods of time. An alternative approach, promoted by the field of ubiquitous computing is to leverage sensors already present in every day devices for estimating health parameters. For example, heart rate can be measured using a smartphone camera by analyzing subtle changes in skin color as heart pumps blood around the body~\cite{shwetak-googlefit,poh2010non}. This technology is now available on billions, of devices\footnote{Google Fit: \url{https://www.google.com/fit/}}. Recent work has presented proof-of-concept measurement of oxygen saturation~\cite{scully2011physiological}, blood pressure~\cite{seismo}, and hemoglobin levels~\cite{wang2016hemaapp} via smartphones.

Existing research work can be broadly divided into two categories: (1) approaches that are developed from first principles 
to imitate an established medical method for measurement or diagnosis~\cite{smartkc, retinoscopy}, and (2) approaches where input (sensor) data and corresponding gold-standard data are collected using a medical grade device and machine learning models are trained to discover a relationship between the input and output \toadd{\cite{ml_1, ml_2}}. In this paper, we focus on the latter category. Although well-intentioned, such data-driven approaches ignore a principled analysis of whether the input data has the necessary information to predict the desired health measure. As a result, numerous human and compute hours are wasted in developing and training deep learning models for prediction tasks which may be ill-posed or not feasible. 

We consider the task of predicting blood pressure (BP) non-invasively. Blood Pressure is the pressure applied on arterial walls as the blood circulates through the body. It depends on multiple factors, including blood volume, blood viscosity, and stiffness of blood vessels.
Abnormally high or low blood pressure can result in heart attack, stroke, diabetes~\cite{heart-attack, diabetes} and thus it is recommended to measure BP frequently.

The methods to measure blood pressure non-invasively can be broadly categorized into two approaches:  i) The pulse transit time (PTT) method~\cite{ptt-1,ptt-2,ptt-3} is a popular, non-invasive technique for measuring blood pressure based on the time delay for a pressure wave to travel between proximal and distal arterial sites. The PTT approach has strong theoretical underpinnings based on the Bramwell-Hill equation\cite{bramwell}, which relates PTT to pulse wave velocity and arterial compliance. The Wesseling model captures the relationship between arterial compliance and blood pressure~\cite{wesseling1993computation}. However, it is important to note that, PTT can change independently of BP due to factors such as aging-induced arteriosclerosis, and smooth muscle contraction. Hence, it needs to be calibrated from time to time. ii) Pulse Wave Analysis (PWA) is a method used to estimate blood pressure (BP) by extracting features from an arterial waveform. This is typically performed using a photoplethysmography (PPG) waveform. PPG is an optical signal obtained by illuminating the skin (common sites are the finger, earlobe, or toe\cite{review_paper}) with an LED and measuring the amount of transmitted, or reflected, light using a photodiode. PPG detects blood volume changes in the microvascular bed of tissue, as the blood volume directly impacts the amount of light transmitted/reflected. Unlike PTT, PWA has weaker theoretical underpinnings as the small arteries interrogated by PPG are viscoelastic\cite{ptt-1}. Calibration is invariably necessary for PWA analysis methods to obtain reasonable results.

In this study, we concentrate on PWA measurement of BP. This method is beneficial because it only requires the use of a single sensor making it a more accessible solution. Predicting BP by analyzing the PPG waveforms is an active area of research~\cite{seismo, ppg2abp, mobicom} and is already used in consumer products\footnote{https://www.samsung.com/global/galaxy/what-is/blood-pressure/}. 
However, we should note that ``\textit{while these methods (PTT and PWA) have been extensively studied and cuff-calibrated devices are now on the
market, there is no compelling proof in the public domain indicating that they can accurately track intra-individual BP changes}''~\cite{review_paper}:. Therefore, although the features extracted from the PPG signal correlate with blood pressure, the signal's adequacy for accurately predicting blood pressure remains unclear.

In this work, we conduct a comprehensive examination of the existing PWA techniques in the literature. Our analysis reveals that a significant portion of these methods are susceptible to one of the four common pitfalls. We analyze these pitfalls in detail in our results section. Later, we present a principled examination of whether the input sensor signal ($x$) (i.e., PPG) can be a good predictor of the output health label ($y$) (i.e., BP). We evaluate if a function $f$ exists, which captures the relationship between $x$ and $y$, such that $y=f(x)$. Moreover, our proposed tool verifies whether the function is well-conditioned or not, i.e., do small changes in $x$ lead to small or large changes in $y$. It is important to ensure that (minor) noise in sensor measurement (which is bound to happen in a real-world setting) does not lead to significant error in the outputs. Our tool is based on information-theoretic notions of \textit{mutual information} and \textit{multi-valued mappings}.
Using our tools, we find that BP prediction using PPG has a high multi-valued mapping factor of 33.2\% and low mutual information of 9.8\%. In comparison, heart rate prediction using PPG, a well-established task, has a very low multi-valued mapping factor of 0.75\% and high mutual information of 87.7\%. This confirms that estimating BP from PPG is an ill-conditioned problem and a more principled approach is needed in future for framing such health measure prediction tasks.

\begin{figure}
\centering
  \includegraphics[width=\textwidth]{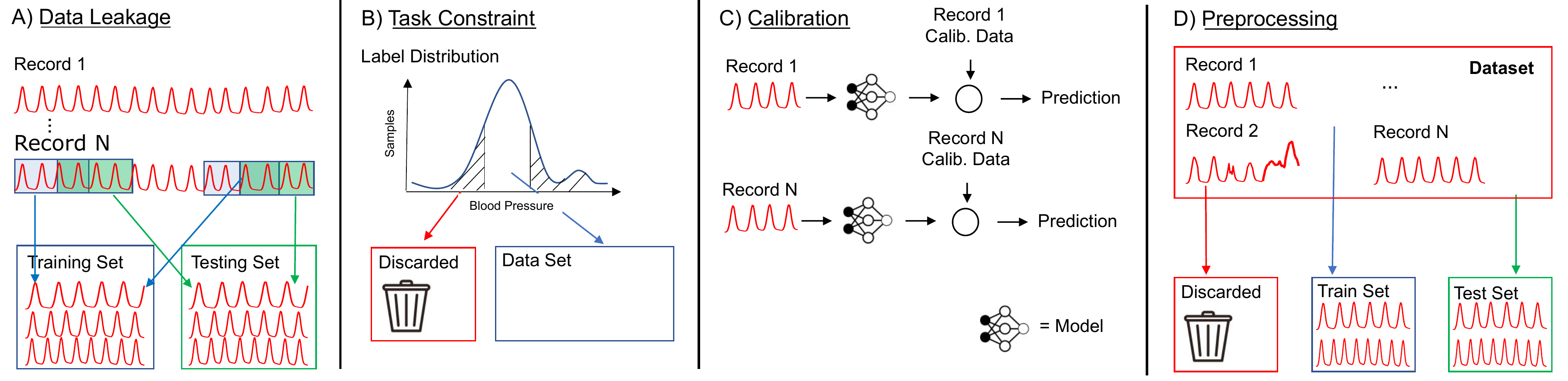}
  \caption{When designing end-to-end machine learning models researchers often use techniques such as: A) providing the model with observations from similar patients, B) constraining the task (e.g., limiting the distribution of labels), C) calibrating models using data from a participant.  When doing so it can often be difficult to identify how these steps impact the integrity of a model, or D) preprocessing to filter out problematic samples (e.g., noisy inputs).}
    \label{fig:teaser}
\end{figure}

%% file: content/3_motivation.tex
\section{Results}
In this section, we present a systematic review of prior work predicting BP via PPG PWA (Figure~\ref{fig:teaser}), followed by a principled analysis using our proposed tools.

\subsection{Review of the Results and Limitations of Prior Work}
\label{sec:prior_work_pitfalls}
To motivate our work, we analyzed recent research~\cite{siamese, brian, mobicom, ppg2abp, spectro_temporal, morph_1, lstm, lstm_2, shallow} which reported results predicting BP via PPG PWA (See Table~\ref{tab:lit_table}). These works relied on the MIMIC~\cite{mimic} dataset (Appendix~\ref{dataset}) containing continuous PPG signals and the corresponding arterial BP values. They evaluated their performance against the AAMI~\cite{AAMI} and/or BHS~\cite{bhs} standards (Appendix~\ref{metric}).
We found that they were prey to some common pitfalls, which resulted in misleading claims and over-optimistic results. For simplicity, we focus on the prediction of Systolic BP (SBP) rather than Diastolic BP (DBP), as SBP has a wider statistical range.

\begin{figure*}[b]
    \includegraphics[width=\textwidth]{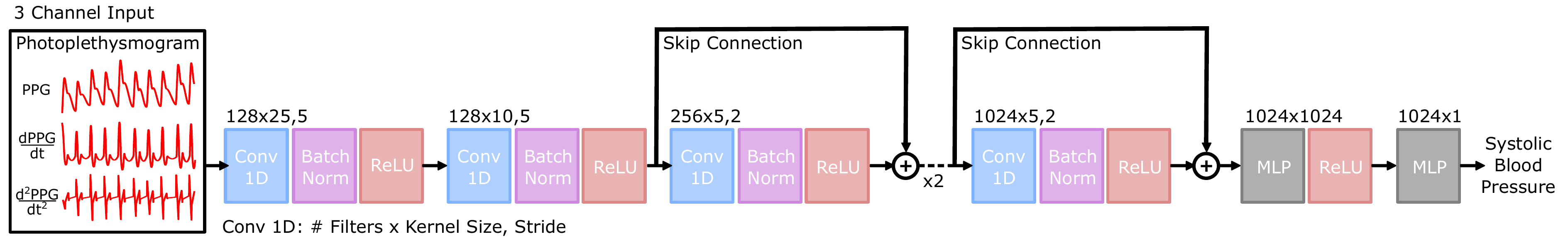}
    \caption{Our reference network, is used to evaluate the impact on performance due to the issues mentioned in Section~\ref{sec:prior_work_pitfalls}. The network has 28M trainable parameters, takes a 3-channel input (PPG, VPG, APG), and outputs the SBP prediction. The model is optimized using a mean squared error loss.} 
    \label{fig:architecture}
\end{figure*}

Before we begin, we should note that not all work (e.g.,~\cite{table-6, table-7, table-8, table-9, lstm}) followed the AAMI/BHS standards accurately. For example, some reported results on a test-set of fewer than 85 subjects. Moreover, despite these works use the same MIMIC dataset, we found a lack of standardization in the train-test data splits and different BP ranges used for evaluation (due to differences in how the data were filtered) across the literature~\cite{siamese, brian}. With the absence of official source code, it was difficult to reproduce prior results and compare different methods. Hence, we trained our own reference deep learning model (Figure~\ref{fig:architecture}), similar to the methods presented in prior research~\cite{siamese, spectro_temporal, morph_1}.
The \textit{reference network} takes a three channel input consisting of the original PPG waveform, along with its first and second derivative, and outputs the predicted SBP value. The model consists of an eight layer residual CNN~\cite{resnet} with 1D convolutions, and is trained using a mean squared error loss. We also explored 2D convolution based CNN models, like Densenet-161~\cite{densenet} and Resnet-101~\cite{resnet},
taking spectrogram of the 1D PPG signal~\cite{siamese} and/or raw waveform as input. Among these, we found that the 1D CNN based architecture performed best.

\begin{figure*}[hbt!]
\centering
    \includegraphics[width=0.67\textwidth]{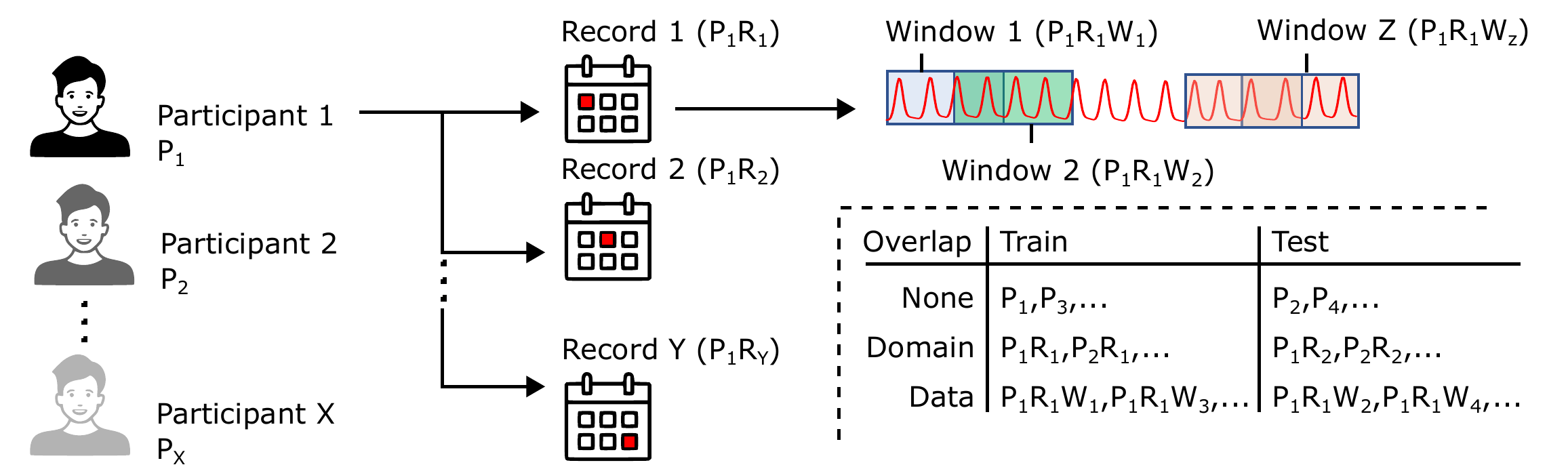}
    \caption{Every participant (P) has multiple data records (R), and each record is divided into multiple overlapping windows (W). Each window forms a data \textit{sample}. In No-Overlap, the train and test data gets split at the participant level, while in Domain-Overlap, the split happens at the record level, and in Data-Overlap, the split happens at the window level.} 
    \label{fig:windows}
\end{figure*}

\subsubsection{Data Leakage}
The goal of any machine learning model is to generalize well to \textit{test} data that will be seen in real-world settings~\cite{pedro_a_few_useful_things_about_ml}. Even with a large training set, it is very unlikely that identical samples to those seen in the training set will appear at test time, thus generalization is crucial. Unfortunately, good performance on a training data set does not always translate to good performance on a test set, as models can \textit{overfit}. This is especially true for modern deep neural networks, which are highly over-parameterized and can easily memorize the training data~\cite{memorization}. Thus, evaluating test performance accurately is an important step in understanding how a model will function in the real world. For this, the test data needs to be pristine, i.e., without any contamination from the training data. Unfortunately, contamination can and does happen in several ways.

We observed two types of overlap between training and testing splits  (Figure~\ref{fig:teaser}A): \emph{data-overlap} and \emph{domain-overlap}. Data-overlap corresponds to overlap of actual samples between the train-test sets. Domain-overlap is more subtle, where although there is no direct overlap of samples, leakage may happen due to similarities in train-test data. In our case, it corresponds to using different records from the \textit{same} patient in both test and train sets (Figure~\ref{fig:windows}).

Here, we consider a particular example from the literature, PPG2ABP~\cite{ppg2abp}, where the authors propose a U-Net based architecture to predict the ABP (Arterial BP) waveform from PPG. They obtain impressive results with a bias of $-1.19$ mmHg and error standard deviation (SD) of 8.01 mmHg\footnote{Note, there is an error in the computation of standard deviation in the PPG2ABP~\cite{ppg2abp} evaluation script. We report the corrected results here.} on the SBP prediction task (Table-\ref{tab:data_leakage_eval}), which is close to the AAMI standard. However, while analyzing their source code, we found both data- and domain-overlaps.

\noindent\textbf{Data-Overlap}: 
The PPG2ABP~\cite{ppg2abp} data processing pipeline divides each PPG \textit{record} ($\sim$6 mins long) into 10-second windows with an overlap of 5 seconds\footnote{github.com/nibtehaz/PPG2ABP/blob/master/codes/data\_processing.py} (Figure~\ref{fig:windows}). Using overlapping windows helps, as it increases the size of training data. However, the problem arises when these 10-second samples are randomly split into train and test sets. Since the overlapping windows are generated \textit{before} the random train-test split, the train and test sets can have samples with the \textit{same} overlapping regions (Figure~\ref{fig:windows}). A deep learning model can memorize values based on these overlapping portions, leading to artificially high accuracy on the test set.

\noindent\textbf{Domain-Overlap}:
Due to the physiological differences between individuals, person-dependent models often outperform person-independent models~\cite{d2015review}. For example, for the BP prediction task, a model can learn the normal range of an individual's BP and leverage that to provide more accurate predictions.
Since the knowledge of an individual's identity can impact a model's accuracy, it is important that the identity of the subject is not leaked (even implicitly) between test and train sets, especially while building person-independent models. Since the PPG signature has been shown to identify an individual~\cite{ppg_biometric}, presence of PPG signals from the same individual in both train and test data can thus leak identity. This turns out to be the case in the PPG2ABP work~\cite{ppg2abp}, as they randomly split PPG records into test and train sets, resulting in different windows from the \textit{same} patient present in both test and train sets (Figure~\ref{fig:windows}).

In order to quantitatively evaluate the impact of data leakage, we compare the performance of the PPG2ABP network on three splits (Figure~\ref{fig:windows}) -- (1) \textit{No-overlap}: The dataset is partitioned at a patient level with a 80-20\% train-test split, (2) \textit{Domain-Overlap}: Each patient has multiple records ($\sim$6 mins long), and these records are randomly split 80-20\% between train-test set, i.e., records from the same patient can be present in both the training and test sets, and
(3) \textit{Data-Overlap}: We use the split provided by PPG2ABP~\cite{ppg2abp} which divides the records into overlapping windows followed by the 80-20\% train-test split. All splits consist of 10-second windows with an overlap of 5-seconds to maintain consistency with the split proposed in PPG2ABP. 
Table-\ref{tab:data_leakage_eval} shows the performance of the PPG2ABP network over the three splits. Domain-overlap significantly increases the accuracy of the PPG2ABP network from a standard deviation of 23.1 to 16.2 mmHg; Data-Overlap further improves the standard deviation to 8.01 mmHg. This analysis clearly shows that leakages, however subtle, can lead to seemingly high but artificial improvements. Note that for all analysis in the rest of this paper, we use the \textit{No-Overlap} split.



\subsubsection{Overconstraining the Task}
\begin{figure}[ht]
\centering
    \includegraphics[width=0.60\textwidth]{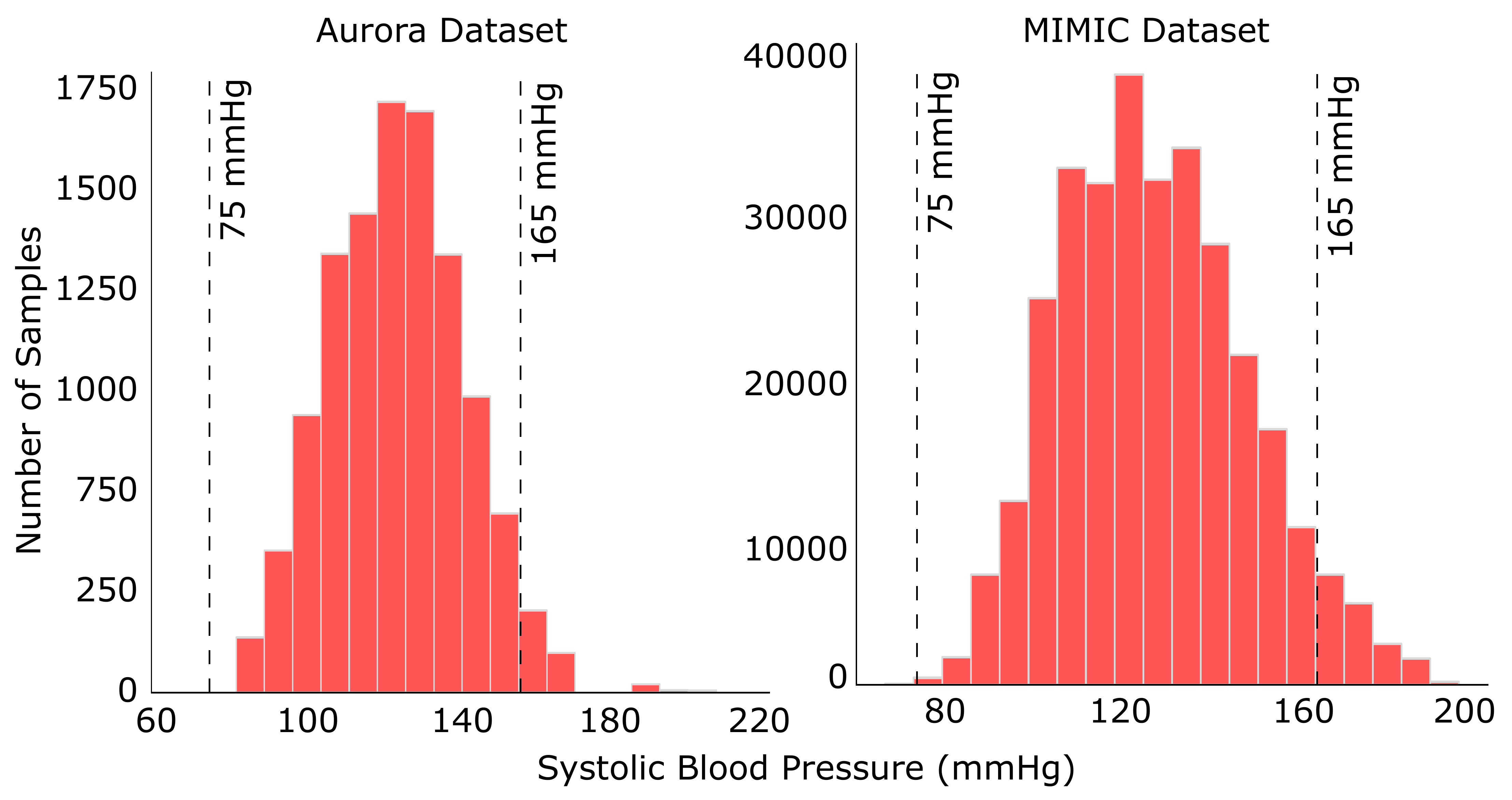}
    \caption{The distribution of systolic BP values in the: (left) Aurora-BP dataset and (right) MIMIC dataset. In the MIMIC dataset, the SBP values lie in the range 65-200 mmHg, however prior works ignore samples with SBP values outside the range of 75-165 mmHg.} 
    \label{fig:dataset_distribution}
    \vspace{-0.3cm}
\end{figure}

Health-related data typically have non-uniform Gaussian distributions, with data density highest near the ``normal'' (or healthy) range, and falling exponentially as we move away from the normal. We observe a similar trend for BP data in both the Aurora-BP~\cite{mieloszyk2022comparison} (Appendix~\ref{dataset}) and MIMIC datasets (see Figure~\ref{fig:dataset_distribution}). While points far from normal are rare, they are often the crucial events (abnormally low or high BP) indicating serious health issues requiring medical attention. 

However, we found that researchers often discard so-called ``outliers''~\cite{siamese, brian, mobicom} (Figure~\ref{fig:teaser}B), arguing that such samples are unlikely or have occurred due to noise in the data collection process. For example, the MIMIC dataset has SBP values ranging between 65-200 mmHg (75-220 mmHg in Aurora-BP), but Schlesinger et al.~\cite{siamese} ignore samples outside the range of 75-165 mmHg, referring to the discarded values as ``improbable''. Similarly, Cao et al.~\cite{mobicom} and Hill et al.~\cite{brian} use a constrained range of 75-150 mmHg, while as per the British Hypertension Society literature, 140-159 mmHg is Grade-1 (mild) hypertension, 160-179 mmHg is Grade-2 (moderate) hypertension, and $\ge$180 is Grade-3 (severe) hypertension~\cite{bhs}.

Constraining the data range has two problems. First, it leads to an incomplete evaluation, as the model is neither trained nor tested on samples from the discarded ranges. Second, since the statistical range of the output is reduced,
this makes the prediction task artificially ``easier'' (i.e., a lower error can be achieved more easily), which may result in promising but misleading results. 
To quantitatively study the impact of constraining data ranges, we conducted an experiment using our reference network with different filtering of data range. Table-\ref{tab:overconstraining} shows the performance of our network when trained with three different SBP ranges--65-200, 75-165 and 75-150 mmHg. Even small restrictions in the output range can lead to a significant (perceived) improvement in accuracy, e.g., reducing the SBP upper limit from 165 to 150 mmHg results in $\sim$11.4\% improvement in the standard deviation. This can be explained as samples at the extremes often results in the highest prediction errors (as models tend to predict closer to the mean of the distribution making predictions on samples with very high or low ground-truth BP values the most inaccurate).


\subsubsection{Unreasonable Calibration}
The relationships between health measures (e.g., PPG and BP) are often person dependent. For example, blood pressure ($bp$) is dependent on the patient's heart rate ($hr$), blood viscosity ($visc$), stiffness of blood vessels ($stif$), etc., i.e., $bp = f(hr, visc, stif, ...)$. While the PPG signal might capture heart rate well, it may not be able to capture viscosity- and stiffness-related information. To solve that, it is common to propose the use of a calibration step, wherein a few PPG samples from each patient along with gold-standard BP values are used to calibrate the function $f$ for that patient (Figure~\ref{fig:teaser}C). The model then learns a calibrated function, $\hat{f}$, for a specific patient, i.e., $bp=\hat{f}(hr)$, where the 
patient-specific parameters ($visc, stif, ...$) are folded into $\hat{f}$.

The literature does not offer a universally effective calibration strategy. Cao et al.'s~\cite{mobicom} method needs to be calibrated every time before a BP prediction to find the optimal fit on the wrist for the watch, while Schlesinger et al.'s~\cite{siamese} model needs to be calibrated once to find the offset value between the model and the true prediction. As blood pressure may not change drastically within minutes (at rest) and significant trends might only be observed over the course of few months due to lifestyle changes or the influence of medication~\cite{bp_change_1}, it becomes important to pay attention to questions like: What is the frequency of re-calibration? 
Is the calibration approach prone to changes in other environmental factors? We believe that the calibration approaches reported in prior work risk over-fitting by memorizing patient-level local temporal characteristics, and that evaluation is incomplete given that they do not evaluate BP prediction over longer time scales.

To understand the influence of calibration, we evaluate the prediction performance under different calibration strategies. \textit{Na\"ive Calibration} simply predicts a constant calibrated value for the entire record. The constant value is computed as the mean of the ground truth values of the first three windows of a record. \textit{Offset Calibration} uses our reference network, but adds an offset to the predicted value. The offset is computed in the calibration step as the difference between the predicted and ground truth BP of the test record's first window. We found the Na\"ive Calibration to perform very well (Table-\ref{tab:3}), with a standard deviation of 8.61 mmHg, close to the AAMI standard. However, predicting a constant BP value for a patient is clearly incorrect. This inconsistency underscores problems with the evaluation methodology. Since typical records in MIMIC are of short time intervals (average length = 6 minutes) compared to the time scales at which BP changes, predicting a constant value gives deceivingly good accuracy. An appropriate evaluation of calibration methods should consider time scales spanning the intended re-calibration duration. For example, if re-calibration is planned every six months, the method should be evaluated with patients tracked over at least a six month time period. 
To demonstrate that calibration systems can quickly deteriorate over time, we analyzed the performance of Offset Calibration as the time from the calibration window increases. Although the method performs well for the first few days, the error rates increase dramatically after that (Figure~\ref{fig:aurora_data}(A)).

\subsubsection{Unrealistic Preprocessing or Filtering}
\begin{figure}[ht!]
\centering
    \includegraphics[width=0.55\textwidth]{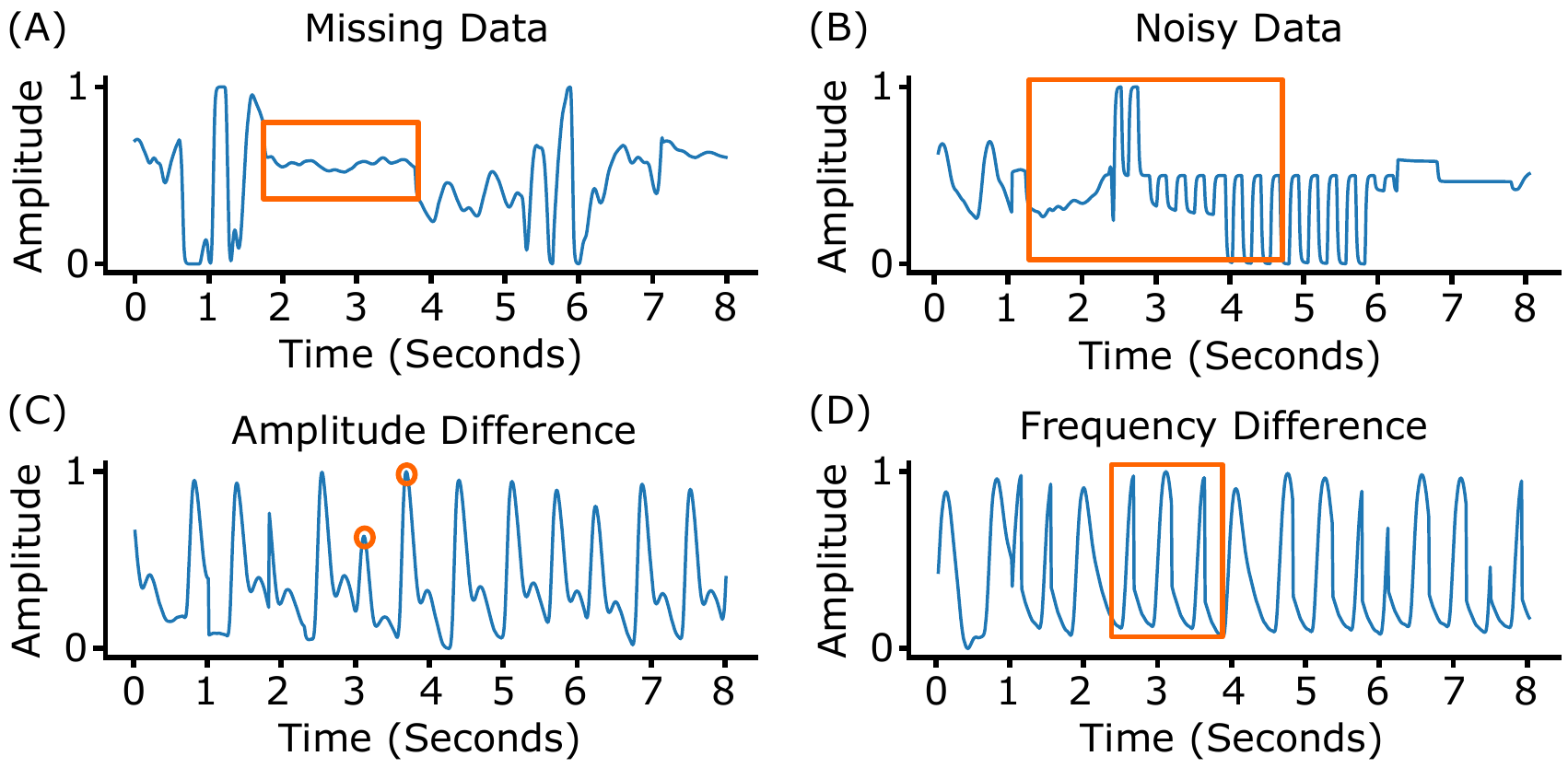}
    \caption{Examples of poor quality photoplethysmography signals from the MIMIC dataset.} 
    \label{fig:filtered_data}
    \vspace{-0.3cm}
\end{figure}

The MIMIC dataset comprises of ICU-patients data, with artifacts due to patient movement, sensor degradation, transmission errors from bedside monitors, and human errors in post-processing data alignment. The impact of these artifacts is visible in both the PPG and ABP waveforms as missing data, noisy data, and sudden changes in amplitude and frequency
(Figure~\ref{fig:filtered_data}). To clean the signal, researchers~\cite{siamese, brian} have used band-pass filters to remove noise in the high frequency ($\ge$16Hz) and low frequency ($\le$0.5Hz) ranges, followed by auto-correlation to filter signals which are not strongly correlated with themselves.
The auto-correlation step removes samples with uneven amplitude and/or frequency.
After cleaning the MIMIC dataset (Figure~\ref{fig:teaser}D), Schlesinger et al.~\cite{siamese} used less than 5\% of the total data for training their neural network, while Hill et al.~\cite{brian} and Slapnicar et al.\cite{spectro_temporal} used less than 10\% of the total MIMIC data. It suggests that ``clean'' data is rare.
Although filtering datasets to remove some noise is often an essential step to train a machine learning model~\cite{pedro_a_few_useful_things_about_ml}, excessive filtering of data can result in overfitting. Models trained on such clean data might achieve high performance on a clean test set; however, they might fail in practice, as it is difficult to obtain such clean signals in a real-world scenario.

\begin{figure}[h!]
\centering
    \includegraphics[width=\textwidth]{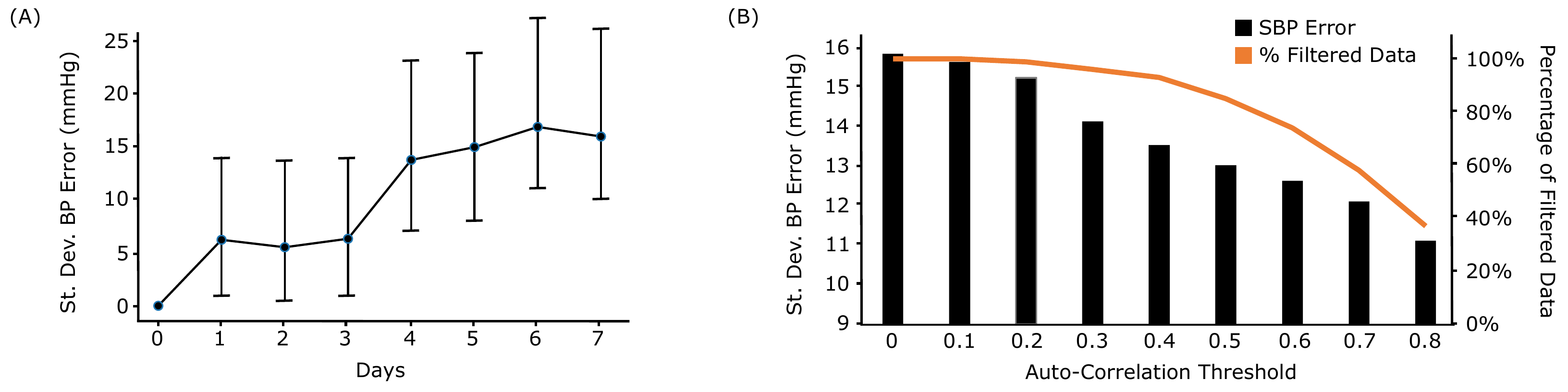}
    \caption{(A) Offset Calibration method's performance falls off quickly after the first few days.
    (B) Performance of our reference network with different auto-correlation thresholds on the MIMIC dataset.}
    \label{fig:aurora_data}
    \vspace{-0.3cm}
\end{figure}

To understand the impact of filtering on a dataset, we measure the performance of our reference network at different auto-correlation thresholds. Figure~\ref{fig:aurora_data}(B) plots the performance of our reference network in predicting SBP and the percentage of filtered data for each auto-correlation threshold. The performance of the network improves by 29.7\% and the dataset size reduces by 63\%, as we increase the auto-correlation threshold from 0 to 0.8.

\subsection{Our Proposed Principled Approach}
We propose and utilize two tools---based on multi-valued mappings and on mutual information (Appendix~\ref{tools})---to estimate if the input signal is a good predictor of the output. Using our proposed tools we performed a principled analysis to study the relationship between PPG and BP. For comparison, we also used our tools on heart rate (HR) and reflected wave arrival time (RWAT) estimation for which it is known that the PPG signal is a strong predictor.

\begin{figure*}
    \includegraphics[width=1\textwidth]{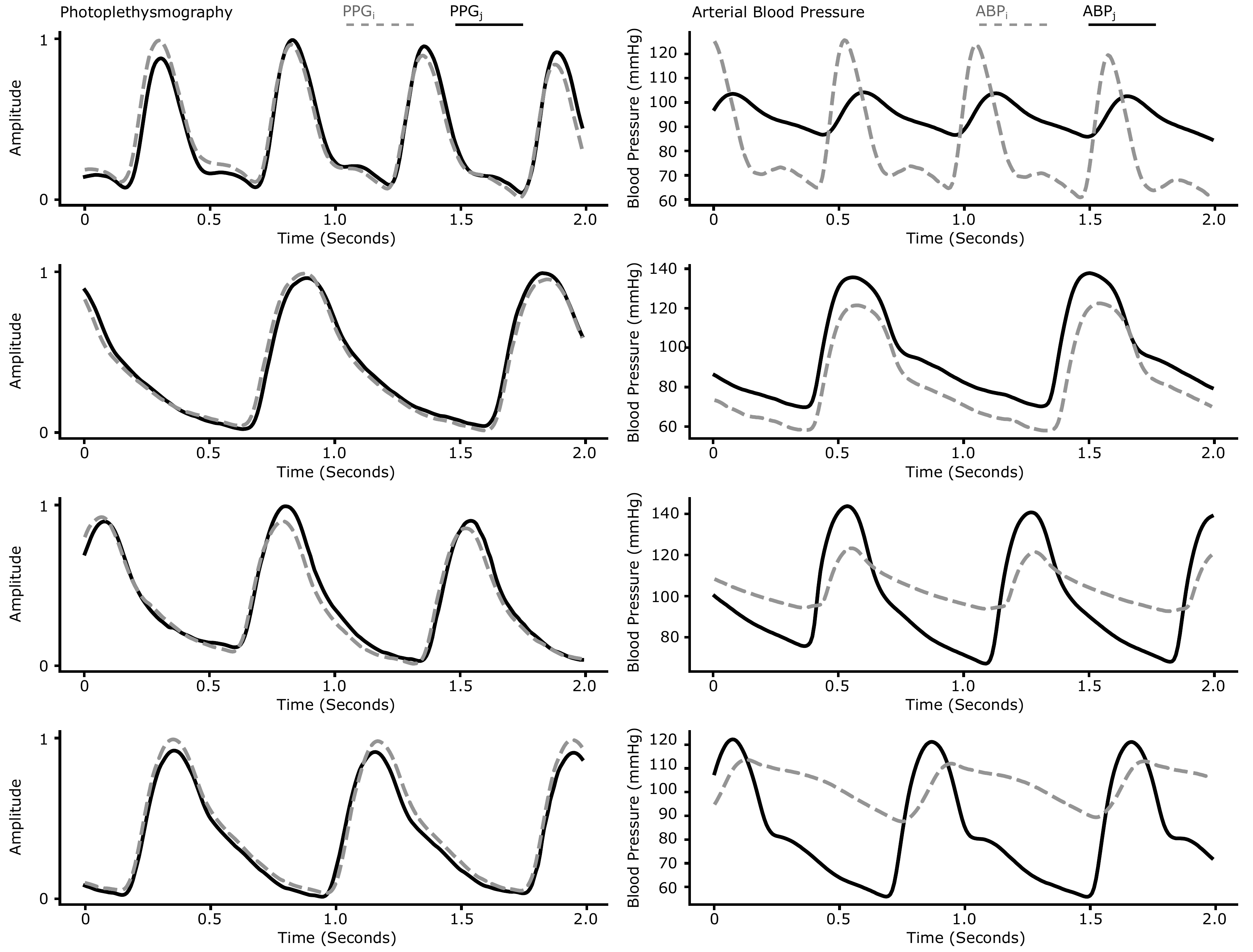}
    \caption{Multi-valued mappings. Examples of PPG waveforms (PPG$_{i}$ and PPG$_{j}$) that are very similar and have corresponding arterial blood pressure waveforms (ABP$_{i}$ and ABP$_{j}$)  that are quite different. This highlights the existence of similar features that map to different targets, which makes the task of blood pressure prediction via PPG pulse wave analysis ill-conditioned.} 
    \label{fig:data_inspektion_results}
\end{figure*}

\textbf{Checking for Multi-Valued Mappings}: We use Algorithm~\ref{algo} to find multi-valued mappings corresponding to data samples that are close in the input space but distant in the output space. As discussed in Section~\ref{sec:mutli-valued-mapping-check}, for computing the distance between two PPG inputs, we first align them using cross-correlation, followed by computing their Euclidean distance. We divide the dataset records into non-overlapping two-second windows and treat them as individual inputs. We set an input distance threshold of $1.0$, which corresponds to a per-time sample threshold of $4e-3$ (each 2s PPG window had 250 samples). For the output, we set thresholds of $8$ mmHg, $8$ bps, and $0.02$s for the BP, HR and RWAT prediction tasks, respectively. 
We found very few multi-valued mappings for the HR and RWAT tasks, while a large number of mappings for the SBP task (Table-\ref{tab:collision_results}). In the MIMIC dataset, for 33.2\% of the 2-second windows, we found another window for the same patient which was close in the input PPG space but had a  significantly different SBP output. When limiting the search to different patients, for 15.0\% of the windows we could still find such matches. This implies that the task of predicting BP from PPG is ill-conditioned. Figure~\ref{fig:data_inspektion_results} shows examples of such multi-valued mappings, with 
highly similar input PPG waveforms but significantly different output arterial BP waveforms. In comparison, for the HR and RWAT tasks, the number of such matches is much smaller at 0.02\% and 0.08\% intra-patient, respectively, suggesting much better conditioning.

\textbf{Evaluating Mutual Information}: For estimating mutual information (MI) between the PPG signal and the target output (BP/HR/RWAT), we use the K-nearest neighbours based approach proposed by Kraskov et al.~\cite{krakow_MI}. We leverage dimensionality reduction to make MI estimation tractable,
using handcrafted and auto-encoder learned feature representations.
We report the mutual information of the input features and target variable, as well as the entropy of the target variable. Note that the target variable's entropy is the maximum achievable mutual information. Thus, the ratio of MI and target variable entropy represents the target information fraction encoded by the input, which we call \textit{Info-Fraction}. We found \textit{Info-Fraction} to be a more intuitive measure than the absolute MI values, and use it to compare the predictive power of PPG across the different tasks.

\begin{figure}[hbt!]
\centering
  \includegraphics[width=0.47\textwidth]{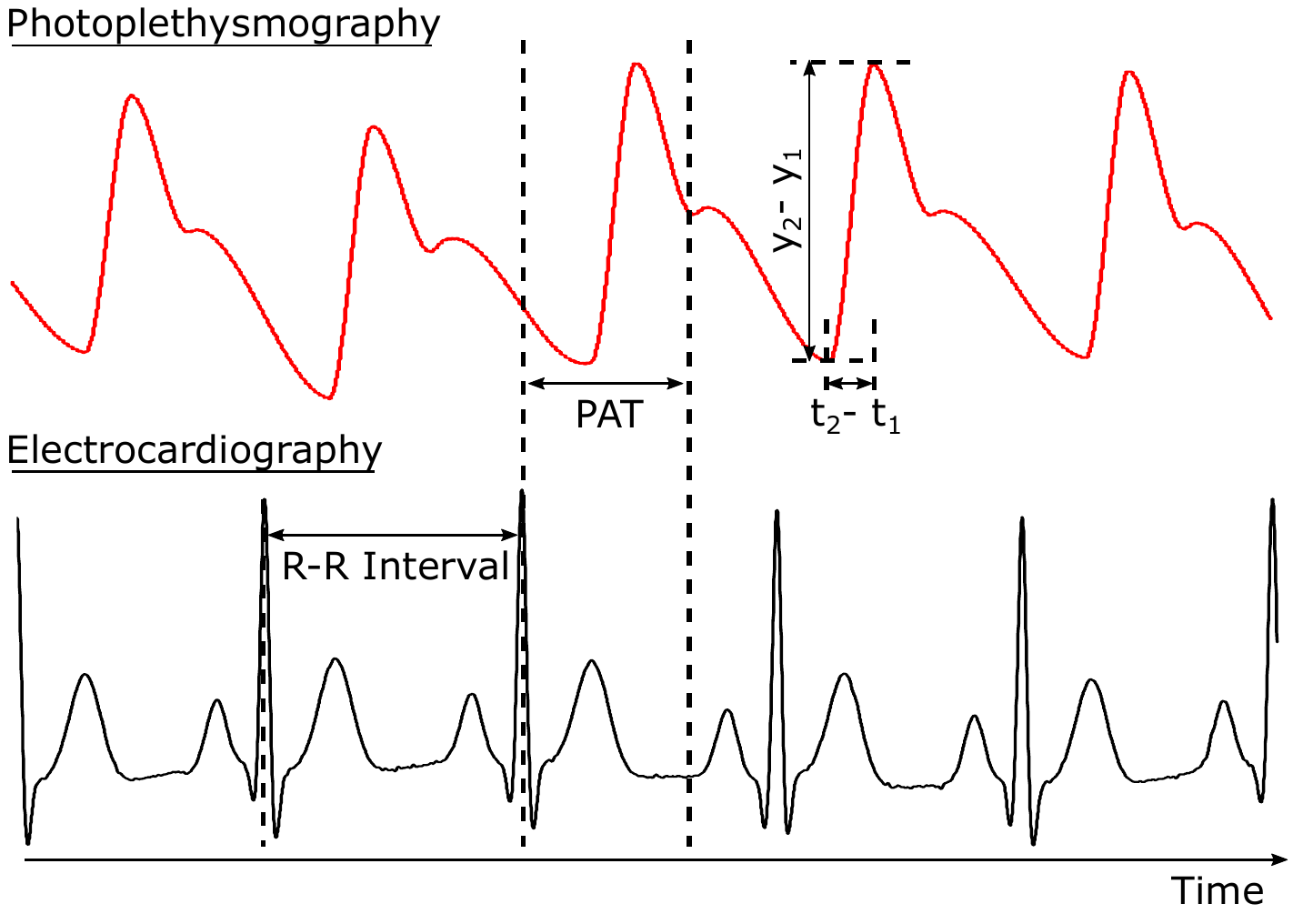}
  \caption{A visual description of the hand-crafted features calculated from the PPG and ECG waveforms. The systolic ramp time ($\frac{dp}{dt}$) is defined as $\frac{y_{2}-y_{1}}{t_{2}-t_{1}}$. } 
    \label{fig:features}
\end{figure}

\textit{Handcrafted Features}: As suggested by Takazawa~\cite{takazawa1993clinical} and Elgendi et al. \cite{elgendi2019use}, we calculate handcrafted features (see Table-\ref{tab:feature_description}) from the PPG waveform
(Figure~\ref{fig:features}).
Due to the absence of a time-aligned ECG waveform in the MIMIC dataset, we extract the relevant handcrafted features only from the PPG waveform. Table-\ref{tab:optical_features_results} presents MI of these individual features with respect to the BP prediction task for both the MIMIC and Aurora-BP datasets, along with MI when all these features are combined and regarded as a single multi-dimensional input. We found that even the combined features set encode a small fraction of the total target entropy. For example, in the MIMIC dataset, the combined features' \textit{Info-Fraction} is just 9.5\%, while heart rate itself contributes an \textit{Info-Fraction} of 4.1\%. Similar observations hold true for the Aurora-BP dataset. This hints that the PPG signal does not have enough information to predict BP in this dataset, and moreover the prediction is highly dependent on the heart rate.

For the Aurora-BP dataset we have the demographic data (age, weight, height) of the subjects, as well as time-aligned PPG and ECG waveforms. This allows us to calculate additional features, e.g., radial Pulse Arrival Time (rPAT) and other derived features~\cite{mieloszyk2022comparison}.
Prior work~\cite{seismo} has used PAT to estimate blood pressure. Moreover, the Aurora-BP dataset has multiple readings for each subject in different positions (e.g., sitting, at rest, and supinated) which helps us add delta features reflecting the difference between features in the two conditions. 
Despite this, we found the entropy results for the Aurora-BP dataset to be similar to the MIMIC dataset, with the handcrafted features able to capture only 9.8\% of the entropy of blood pressure (Table-\ref{tab:feature_results}). On the other hand, for the HR and RWAT prediction tasks, the handcrafted features captured 87.7\% and 64.6\% entropy, respectively.\footnote{The ground truth for HR is derived from the ECG sensor data and RWAT from the tonometric sensor data.}
This further strengthens our finding that the PPG signal even with additional information from the ECG waveform has limited information to predict BP.

\textit{Auto-encoder Features}:
As an alternative to handcrafted features, we train an auto-encoder on the raw PPG waveform to obtain a set of low dimensional features. We use a five layer multi-layer perceptron (MLP) auto-encoder with ReLU activation and a bottleneck layer of 20 neurons. The model was trained with Adam optimizer (learning rate of 0.001) and a mean-squared error loss (with a stopping point when the loss saturated at $<$0.1).
Training time on a single NVIDIA P100 was under an hour. Table-\ref{tab:autoencoder_results} shows MI of the combined bottleneck features with respect to the BP, HR and RWAT prediction tasks. Although the auto-encoder features are more comprehensive and have higher MI compared to the hand-crafted features, the \textit{Info-Fraction} for BP prediction (12.9\% for MIMIC and 8.7\% for Aurora-BP) is still much lower compared to that for HR (92.2\% for MIMIC and 93.1\% for Aurora-BP) and RWAT (70.1\% for Aurora-BP) prediction tasks.


%% file: content/7_discussion.tex
\section{Conclusion}
Our results reveal that BP prediction via pulse wave analysis of the PPG signal is still an unsolved task and far from the acceptable AAMI and BHS standards. By performing a systematic review and accompanying experiments we have found several flaws in the prior work that have lead to seemingly over-optimistic results. Examples of problems include using data splits that leak information from test samples into the training set, heavy constraints on the task that remove challenging samples and reduce the variance of target values substantially, calibration methods that seem to be practically problematic,
and unreasonable preprocessing that cleans the data to an unrealistic extent such that any noise is unacceptable. Our results do not prove that blood pressure estimation from PPG PWA is impossible; however, they do suggest the task is very ill-conditioned, and that fairly evaluating performance is non-trivial. We have proposed a set of tools based on multi-valued mapping and mutual information to check if an input signal is a good predictor of the desired output. The multi-valued mapping checker allows us to find samples close in input space but far in the output space. We found many such samples in both MIMIC and Aurora-BP datasets.
Searching for multi-valued mappings was trivial once appropriate distance metric and thresholds were defined, and qualitative and quantitative results show that almost identical PPG waveforms can have very different blood pressure waveforms. Next, we looked at the entropy of the features by computing mutual information. MI was extremely low for both hand-crafted and learned auto-encoder features. In comparison, heart rate and RWAT prediction tasks from PPG PWA have much lower multi-valued mapping factors and a much higher mutual information. Combining pitfalls in prior work and analysis from our proposed tools confirm that predicting BP from PPG signal is an ill-conditioned task.

%% file: content/X_Appendix.tex
\begin{table}[hbt!]
\begin{center}
\caption{Performance of PPG2ABP~\cite{ppg2abp} on different test-train splits with varying degrees of dataset overlap. Even subtle leakages can result in large (but artificial) accuracy improvements.}
\begin{tabular}{ l c c } 
  \toprule[1.5pt]
  &\multicolumn{2}{c}{PPG2ABP~\cite{ppg2abp}} \\
  Data Split & Bias (mmHg) & SD (mmHg) \\
  \midrule
  No-Overlap & 1.11 & 23.1 \\ 
  Domain-Overlap & 5.12 & 16.2   \\ 
  Data-Overlap & -1.19 & 8.01 \\ 
  \bottomrule[1.5pt]
\end{tabular}
\label{tab:data_leakage_eval}
\end{center}
\end{table}

\begin{table}[hbt!]
\begin{center}
\caption{Performance of reference network on different SBP ranges on the MIMIC dataset. Constraining the data range can result in significant (but artificial) accuracy improvements.}
\begin{tabular}{ l c c } 
  \toprule[1.5pt]
  SBP Range (mmHg) & Bias (mmHg) & SD (mmHg)  \\
  \midrule
  65-200 &  -3.45 & 15.8 \\ 
  75-165 & -4.59 & 14.0 \\ 
  75-150 & -4.42 & 12.4 \\ 
  \bottomrule[1.5pt]
\end{tabular}
\label{tab:overconstraining}
\end{center}
\end{table}

\begin{table}[hbt!]
\begin{center}
\caption{Performance of different calibration-methods on the MIMIC dataset. The incorrect Na\"ive calibration methods performs very well, underscoring problems with the evaluation methodology.}
\begin{tabular}{ r c c} 
  \toprule[1.5pt]
  Method & Bias (mmHg) & SD (mmHg)\\
  \midrule
  Na\"ive Calibration & 0.79 & 8.61  \\ 
  Offset Calibration & 0.38 & 9.82  \\
  No Calibration & 0.28 & 10.9  \\ 
  \bottomrule[1.5pt]
\end{tabular}
\label{tab:3}
\end{center}
\end{table}

\begin{table}[hbt!]
\begin{center}
\caption{Multi-valued mapping matches for the BP, HR and RWAT predictions tasks. For the BP task, there was high match rate for both within the same patient records and across patients, suggesting an ill-conditioned problem. For the HR and RWAT tasks, the matches were much lower. Ground truth for RWAT is only available for the Aurora-BP dataset.}
\begin{tabular}{ r c c c c }
  \toprule[1.5pt]
  &\multicolumn{2}{c}{MIMIC} & \multicolumn{2}{c}{Aurora-BP} \\
  Task & Intra-Patient & Inter-Patient & Intra-Patient & Inter-Patient \\
  \midrule
  SBP  & 33.2\%  & 15.0\% & 13.9\% & 16.2\% \\
  HR   & 0.75\%  & 2.10\% & 0.02\% & 0.89\% \\
  RWAT & -  & -           & 0.08\% & 4.78\% \\

  \bottomrule[1.5pt]
\end{tabular}
\label{tab:collision_results}
\end{center}
\end{table}

\begin{table}[hbt!]
\begin{center}
\caption{Descriptions of the handcrafted features used for the Mutual Information analyses.}
\renewcommand{\arraystretch}{2}
\begin{tabular}{p{1.5cm} p{10cm}} 
  \toprule[1.5pt]
  Feature & Description\\
  \midrule
   Heart Rate (HR) & Measurement of the number of pulsations of the heart in a minute. Calculated as the inverse of median time between each heart beat. The PPG signal was used for MIMIC (because time alignment with the ECG signal was not precise), the ECG signal was used for Aurora-BP.\\ 
   Heart Rate Variability (HRV) & Measurement of the variation in time between each heart beat. Calculated as the mean of standard deviations of normal-normal (NN) intervals (SDNN). \\
   Quality & Measurement of the quality of the PPG signal. A heuristic based algorithm that takes the signal-to-noise ratio, artifacts, consistency between the pulses in a window into consideration and computes a normalized score between 0 to 1.\\
   $\frac{dp}{dt}$ & Measurement of the mean of the systolic rise times normalized with respect to the duration of each beat in the PPG signal. \\
   rPAT & Measurement of the delay between the R-peak in the ECG signal and systolic peak of the PPG signal. This can only be computed for Aurora-BP due to imprecise synchronization in MIMIC.\\
   Inv. PAT & 1/rPAT.\\
   $\Delta$ Feature & Measured as the difference between the calculated value and baseline value. A baseline value of each feature for all patients is computed in Aurora-BP (not available for MIMIC). \\
   std.Feature & Measures the fluctuation of a feature across a fixed time period. \\
  \bottomrule
\end{tabular}
\label{tab:feature_description}
\end{center}
\end{table} 

\begin{table}
\begin{center}
\caption{Mutual Information of PPG optical features in the BP prediction task. Even all features combined have a small \textit{Info-Fraction}, and most of that is captured by the heart rate feature alone.}
\begin{tabular}{c c c}  
\toprule
& \multicolumn{2}{c}{Mutual Information (bits)}\\
Optical Features & MIMIC & Aurora-BP \\ 
\midrule
HR & 0.120 & 0.103\\ 
HRV & 0.070 & 0.054\\
Quality & 0.070 & 0.112\\
$\frac{dp}{dt}$ & 0.013 & 0.064 \\ 
\hdashline
Combined & 0.280 & 0.240 \\ \hline 
Entropy & 2.930 & 3.680\\
\textit{Info-Fraction} (Combined) & 9.5\% & 6.5\% \\
\textit{Info-Fraction} (HR) & 4.1\% & 2.8\% \\
\bottomrule[1.5pt]
\end{tabular}
\label{tab:optical_features_results}
\end{center}
\end{table}

\begin{table}[hbt!]
\begin{center}
\caption{Mutual Information of patient demographic data, PPG optical features and features derived using ECG, for the Aurora-BP dataset~\cite{mieloszyk2022comparison}.
While all features combined have an \textit{Info-Fraction} of just 9.8\% for the SBP prediction task, they encode a lot more information for the HR prediction (87.7\%) and RWAT prediction (64.6\%) tasks.}
\begin{tabular}{r c c c }
\toprule[1.5pt]
&\multicolumn{3}{c}{Mutual Information} \\ 
Feature & SBP (bits) & HR (bits) & RWAT (bits)\\ 
\midrule
Age & 0.026 & 0.015 & 0.000 \\
Weight & 0.024 & 0.024  & 0.000  \\
Height & 0.007 & 0.000 & 0.000\\
HR & 0.130 & 2.000 & 0.658 \\ 
std HR & 0.009 & 1.230 & 0.509\\
rPAT & 0.200 & 0.220 & 0.116\\ 
HRV & 0.170 & 0.295 & 0.131\\
Inv. PAT & 0.070 & 0.132 & 0.065\\
Quality & 0.100 & 0.080 & 0.000\\
$\frac{dp}{dt}$ & 0.016 & 0.476 & 0.584\\ 
std $\frac{dp}{dt}$ & 0.009 & 0.232 & 0.252\\
$\Delta$ rPAT & 0.160 & 0.165 & 0.074\\
$\Delta$ Inv. PAT & 0.060 & 0.072 & 0.063\\ 
$\Delta$ $\frac{dp}{dt}$ & 0.014 & 0.131 & 0.139\\
$\Delta$ HRV & 0.170 & 0.130 & 0.016 \\ 
$\Delta$ HR & 0.025 & 0.242 & 0.234\\ 
$\Delta$ Quality & 0.070 & 0.060 & 0.001\\
\hdashline
Combined &  0.364 &  3.240 & 1.650\\ \hline 
Entropy & 3.680 & 3.650 & 2.540\\
\textit{Info-Fraction} (Combined) &  9.89\% & 88.8\% &  65.0\%\\
\bottomrule[1.5pt]
\end{tabular}
\label{tab:feature_results}
\end{center}
\end{table}

\begin{table}
\begin{center}
\caption{Mutual information of auto-encoder features. The same trend of Table-\ref{tab:feature_results} holds. While the SBP task has low \textit{Info-Fraction}, the features encode a lot more information for the HR and RWAT tasks. Ground truth for RWAT is only available for the Aurora-BP dataset~\cite{mieloszyk2022comparison}.}
\begin{tabular}{ c c c c c c c}  
\toprule[1.5pt]
& \multicolumn{3}{c}{MIMIC} & \multicolumn{3}{c}{Aurora-BP} \\
Task & MI & Entropy & \textit{Info-Fraction} & MI & Entropy & \textit{Info-Fraction} \\ 
\midrule
SBP  & 0.38 & 2.93 & 12.9\%  & 0.32 & 3.68 & 8.70\% \\
HR   & 2.60 & 2.82 & 92.2\%  & 3.40 & 3.65 & 93.1\% \\
RWAT  & - & - & -             & 1.78 & 2.54 & 70.1\% \\
\bottomrule[1.5pt]
\end{tabular}
\label{tab:autoencoder_results}
\end{center}
\end{table}

\begin{table}
\footnotesize
\renewcommand*{\arraystretch}{1.2}
\caption{
The table summarizes the limitations of previous research and indicates whether the study exhibits specific pitfalls. The pitfalls are categorized into four categories:
a) Data-split: Domain Overlap (denoted as D.O), Data Overlap (denoted as C.O), or Small test set (denoted as S.T).
b) Over-constraining: SBP values and standard deviation (if provided)
c) Unrealistic Pre-Processing: \% of remaining dataset after pre-processing (if provided)
d) Calibration: Correctly employed and justified for longer periods.
The columns denote the presence or absence of each limitation, with Y (yes) indicating the study has the limitation, N (no) indicating it does not have that limitation, U (unknown) indicating that there is not enough information available, and "-" indicating that the research is not applicable to the pitfall. For additional information, please see Section 2.1.
}
\noindent\makebox[\textwidth]{
\begin{tabular}{p{1.5cm}p{1.5cm}p{1.5cm}p{1.5cm}p{1.5cm}p{1.5cm}p{1.cm}} 
\toprule[1.5pt]
Method & Dataset & Results (SBP) & Data-split & Over-constraining & Unrealistic Pre-Proc. & Calibration\\
\hline \hline
BiGRU Attention\cite{table-1}  & MIMIC-II & MAE=2.58 SD=3.35 & U & N \newline    SD=14.1 & N \newline 
    $\sim$10\% & \textemdash \\
\hline
AdaBoost\cite{table-2}  & MIMIC-II & ME=0.09 MAE=8.22 SD=10.38 & N \newline D.O & Y & Y & \textemdash \\
\hline
ANN\cite{table-3} & MIMIC-II & MAE=3.21 RMSE=4.23 & U & Y & N \newline $\sim$75\% & \textemdash \\ 
\hline
LSTM\cite{table-4} & MIMIC-II & MAE=3.23 STD=4.75 & U & U & Y & \textemdash \\
\hline
Ensemble 1-D CNN 2-D CNN \cite{spectro_temporal}& MIMIC-III & MAE=9.43 & Y\newline S.T & Y & N \newline $\sim$1.7\% & \textemdash  \\ 
\hline
ANN \cite{table-6} & MIMIC & MAE=4.02 SD=2.79 & U \newline S.T & U & Y & \textemdash \\
\hline
Regression\cite{table-7} & Custom Dataset & MAE=6.90 SD=9.00 & Y\newline S.T & Y & Y & \textemdash \\
\hline
SVR\cite{table-8} & Queensland & ME=11.6 SD=8.20 & Y\newline S.T & Y & Y & \textemdash \\
\hline
Regression\cite{table-9} & Custom Dataset & MAE=3.90 SD=5.37 & Y\newline S.T & N & Y & \textemdash \\
\hline
ANN\cite{table-10} & MIMIC-II & ME=0.16 MAE=4.47 SD=6.85 & N \newline C.O & U & Y & \textemdash \\
\hline
SVR \cite{table-11} & Custom Dataset & ME=5.10 SD=4.30 & Y \newline S.T & N \newline SD=11.9 & Y & \textemdash \\
  \hline
SVR\cite{table-12} & Queensland & MAE=4.76 SD=7.68 & N \newline D.O \newline S.T & N & Y & \textemdash \\
\hline
ANN\cite{table-14}  & MIMIC & MAE=3.80 SD=3.46 & U\newline S.T & N & N & \textemdash \\
\hline
Regression\cite{table-15} & MIMIC & MAE=4.90 SD=6.59 & N \newline D.O \newline S.T & Y & Y & N \\
\hline
LSTM-CNN\cite{table-16}  & MIMIC-II & ME=1.55 SD=5.41 & U \newline S.T & N & N\newline $\sim$15\% & \textemdash \\
\hline
AdaBoost\cite{table-17}  & MIMIC-II & ME= -0.05 SD=8.90 & N \newline D.O & Y & N\newline $\sim$20\% & \textemdash \\
\hline
U-Net\cite{ppg2abp} & MIMIC-II & ME=-1.58 SD=8.61 & N\newline C.O & Y & Y & \textemdash \\
\hline
CNN Siamese\cite{siamese} & MIMIC-II & MAE=5.95 SD=6.90 [Calib] & Y & N & N\newline $\sim$5\% & N \\
\hline
U-Net\cite{brian}  & MIMIC-II 
& ME=4.30 SD=6.50
ME=2.30 SD=5.60
& Y & N\newline SD=13.5 & Y & \textemdash \\
\hline
1-D CNN\cite{morph_2}  & Custom Dataset & SD=11.4 & N \newline D.O \newline S.T & N\newline SD=16 & Y & \textemdash \\
\hline
LSTM\cite{lstm_2} & MIMIC-II & ME=4.05 SD=4.60  & N\newline D.O\newline S.T & N & N\newline $\sim$50\% & \textemdash \\
\end{tabular}
}
\label{tab:lit_table}
\end{table}

\appendix
\label{appendix}

\input{content/2_related_work}
\input{content/4_solutions}

\section{Data Availability Statement}
All the data used in this work is publicly available.  The MIMIC~\cite{mimic_original} and Aurora-BP~\cite{mieloszyk2022comparison} datasets can be accessed by researchers after completing the necessary steps stated by the creators of those datasets.

\input{content/6_results}

%% file: content/2_related_work.tex
\section{Related Work} 

The gold-standard for blood pressure measurement, used in Intensive Care Units and Operating Theatres, requires an invasive procedure that involves inserting a cannula needle into an artery. The cannula needle is connected to a transducer which converts the pulse signal to the arterial pressure waveform, providing continuous pulse-level BP measurements. Such invasive measurement is not feasible outside of a hospital setting, therefore two alternative cuff-based non-invasive procedures---auscultatory and oscillometry methods---are widely used~\cite{bonnafoux1996auscultatory}. However, these methods do not provide continuous measurement, hence researchers~\cite{seismo, da2010ear, holz2017glabella} have been actively working on developing novel methods to accurately estimate blood pressure in a non-invasive continuous manner. A majority of the proposed methods involve calculating Pulse Transit Time (PTT) which is inversely correlated to BP. PTT is defined as the time taken by a pulse to travel between two arterial sites---one measured using PPG and the other captured from a different sensor. E.g., Ding et al.~\cite{ding2015continuous} captured ECG, He et al.~\cite{da2010ear} used Ballistocardiogram from the ear, Holz and Wang~\cite{holz2017glabella} collected accelerometer signals from head, and Wang et al.~\cite{seismo} captured accelerometer signals using a smartphone pressed to the chest.

Considering the ease and accessibility of accurately measuring heart rate and heart rate variability via PPG captured from a smartphone or wearable, a natural extension is to attempt to calculate blood pressure solely by analyzing the PPG pulse wave. Recent works~\cite{siamese, mobicom, brian, lstm, lstm_2} have explored and published promising results for the BP prediction task from PPG pulse wave analysis.
These proposed methods involve building data-driven regression models to learn meaningful features by leveraging the availability of large PPG-BP labelled datasets (MIMIC~\cite{mimic}). For example, Schlesinger et al.~\cite{siamese} predicts BP using Convolution Neural Networks (CNN) trained on a frequency domain representation of the PPG signal and uses siamese logic to calibrate BP predictions at run-time, Tazarv and Levorato~\cite{lstm} used a Long Short-Term Memory (LSTM) network with the PPG waveform as input, and Slapnicar et al.~\cite{spectro_temporal} proposed an ensemble network of 1-D CNNs and LSTMs on the raw and first two derivatives of the PPG signal. Some recent works~\cite{brian, ppg2abp} have proposed an extension to the prior work by predicting the full Arterial Blood Pressure (ABP) waveform from the PPG signal using U-Net based architectures.

%% file: content/4_solutions.tex
\section{Methods}\label{tools}
We propose two tools---based on multi-valued mappings and on mutual information---to estimate if the input to a model is a good predictor of the output.

\subsection{Multi-valued Mapping Check}
\label{sec:mutli-valued-mapping-check}
If the input sensor signal ($x$) is a good predictor of an output health labels ($y$), it means there exists a function $f$, such that $y=f(x)$. Moreover, the function $f$ should be well-conditioned, i.e., small changes in $x$ should not lead to large changes in $y$. This is important to ensure that small amounts of noise in the sensor measurement (which are bound to happen in a real-world setting) do not lead to significant errors in the output. To test whether a task is well-conditioned, we propose searching for multi-valued mappings using Algorithm~\ref{algo}. 
\begin{algorithm}[H]
	\caption{Multi-valued Mapping Search} 
	\label{algo}
	\begin{algorithmic}[1]
	    \State multivalued\_mappings = \{\}
	    \State ppg[i], sbp[i] = ppg wave at index i, sbp value at index i 
	    \State N = size of ppg and sbp arrays
	    \State dist(x$_{1}$, x$_{2}$) = function to calculate distance between ppg wave x$_{1}$ and x$_{2}$
	    \State t$_{i}$ = threshold for input space
	    \State t$_{o}$ = threshold for output space
		\For {$i=1,2,\ldots, N$}
			\For {$j=1,2,\ldots,N$}
			    \If {$dist$(ppg[i], ppg[j]) $\le$ t$_{i}$ 
                    \textbf{and} $abs$(sbp[i]-sbp[j]) $\ge$ t$_{o}$}
			        \State multivalued\_mappings.add([i,j])
	        	\EndIf
			\EndFor
		\EndFor
	\end{algorithmic} 
\end{algorithm} 
\vspace{-0.1cm}
Our multi-valued mapping algorithm searches for samples that are close in the input space but distant in the output space. If the algorithm is able to find such mappings, it means that the function $f$ either does not exist, or is at best ill-conditioned.
Algorithm~\ref{algo} has two key components: a distance function for comparing the input samples and an optimal threshold for filtering multi-valued mappings.

\noindent\textbf{Distance Function}: Searching for multi-valued mappings in a dataset requires a metric to quantify the distance between the input samples. However, choosing the right distance function is not always obvious, and one needs to be careful about the implicit assumptions in any given metric. For example, cross-correlation, dynamic time warping (DTW)~\cite{bellman1959adaptive}, and Euclidean distance are ways to measure the distance between two time-series/waveforms, and each has specific characteristics---cross-correlation is phase invariant, DTW is scale invariant in the time dimension, and Euclidean distance is translation invariant. For cross-correlation, a sliding window dot-product of the two input data series is computed to find the point where the similarity is maximized;
DTW computes an optimal match by reducing the minimum-edit distance between the two series; Euclidean distance measures the similarity between the two
data series using the L2 distance.

Ideally, the distance function should align well with the task requirements. Among the three distance functions, DTW makes the similarity metric invariant with respect to time scale. However, it is known that BP has a direct dependency on heart rate, which in turn is determined by the periodicity of the PPG waves. Thus, the time scale invariance property of DTW will result in information loss for this task, making it a bad choice as a distance function. Euclidean distance used in isolation is not a good choice either, as even the same PPG signals slightly shifted in time can result in a high Euclidean distance value. Since the relationship between PPG and BP should not change with small shifts of the PPG signal forward or backward in time, such a distance metric is not suitable. Therefore, cross-correlation is ideal to create an appropriate distance metric. Although cross-correlation based distance metric worked well in our experiments, we found that aligning PPG signals via cross-correlation followed by computing Euclidean distance between the aligned signals looked logical. We used this distance measure for all our experiments.

\begin{figure}[hbt!]
\centering
  \includegraphics[width=0.55\textwidth]{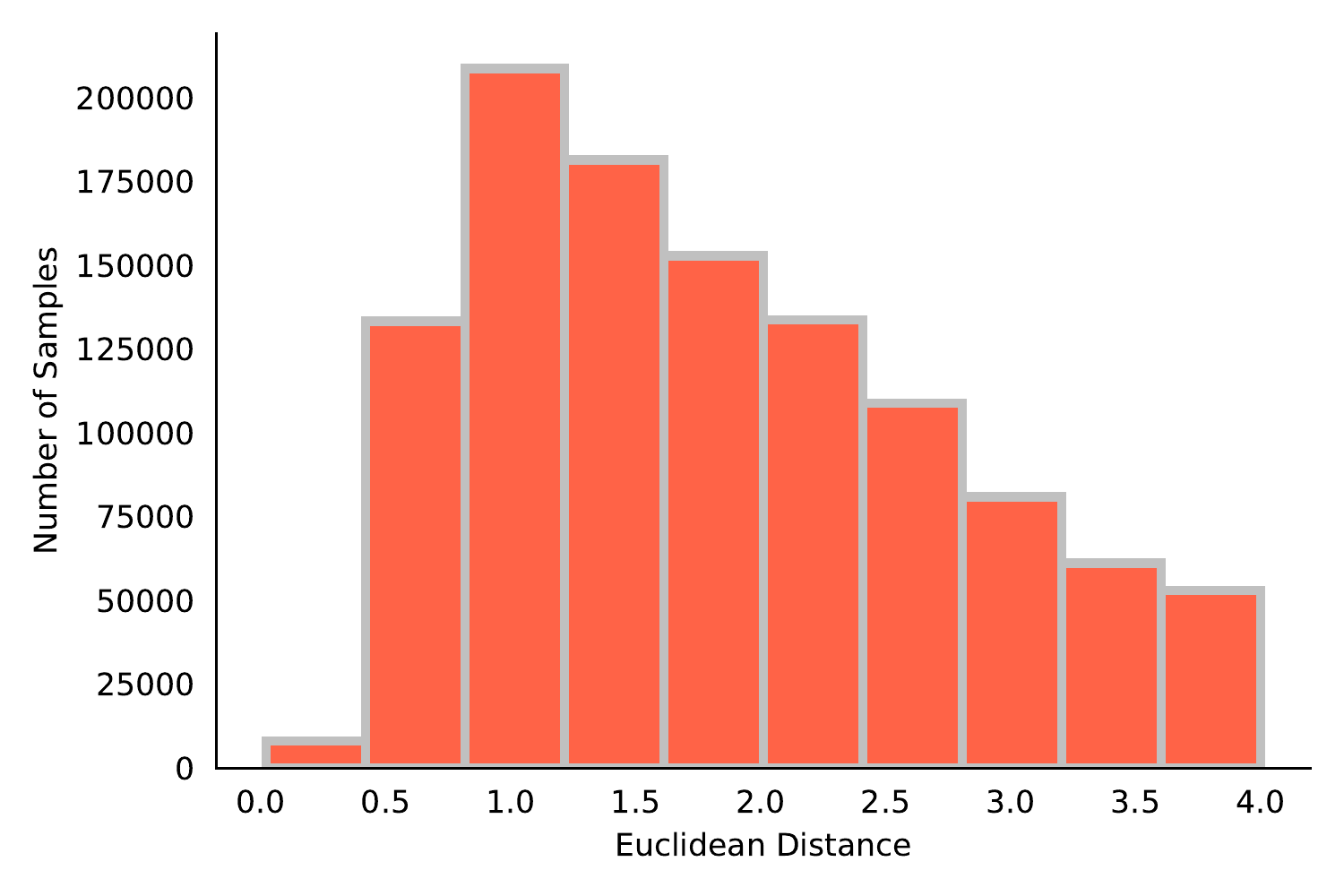}
  \caption{The distribution of Euclidean Distances between pairs of aligned consecutive PPG waves.}
    \label{fig:threshold}
\end{figure}

\noindent\textbf{Optimal Threshold}:
After choosing the appropriate distance function, we need to identify an optimal distance threshold, below which two signals can be considered ``equal''. However, it is not straightforward to find such a threshold. If the threshold is very generous (i.e., high), we will end up selecting distant input signals as equal, and get misleading multi-valued mappings. On the other hand, if the threshold is too strict (i.e., low), we may not find any multi-valued mappings even for ill-conditioned functions, as the chances of two input signals being identical, especially in the presence of noise, is very small. 
To identify the optimal threshold for filtering multi-valued mappings, we calculate the Euclidean distance between two consecutive aligned PPG waves, each 2 seconds in duration. This interval was chosen as it represents an ideal time frame in which the signal remains consistent. Ideally, the difference between 2 consecutive PPG waves should account for an irreducible error, and this can be used as a threshold for filtering multi-valued mappings. 
Figure-\ref{fig:threshold} illustrates the results of this analysis, which indicates that a majority of the PPG waves exhibit a Euclidean distance of $\le$ 1, which led us to choose 1.0 as the threshold for our experiment.

Note that our multi-valued mapping check is a one-way method, i.e., if we are able to find multi-valued mappings, it implies an ill-conditioned $f$; however not finding multi-valued mappings does not guarantee existence of a well-defined $f$. This is because Algorithm~\ref{algo} may fail to find signals close in the input space due to sparsity of the dataset. The mutual information check discussed next provides a complimentary method.

\subsection{Mutual Information Check}
\label{sec:mutual_information_check}
Mutual Information (MI) is an information theoretical measure of the dependence between two random variables $X$ and $Y$, defined as:
\begin{equation} \label{eq3}
\begin{split}
I(X; Y) & = H(X) - H(X \vert Y)  \\
 & = H(Y) - H(Y \vert X), \\
\end{split}
\end{equation}
where $H$ is the Shannon entropy function ($H(X) = -\sum_{i} p(x_i) log(p(x_i))$\footnote{For continuous analog data, it is computed via limiting density of discrete points (LDDP)~\cite{lldp}}). The marginal entropies $H(X)$ and $H(Y)$ represent the amount of information needed to describe the outcome of the random variable. This is same as the uncertainty of the random variable. $H(X \vert Y)$ and $H(Y \vert X)$ are conditional entropies, and denote the amount of information needed to describe the outcome of one random variable when the value of the other variable is known. This can also be thought of as the amount of uncertainty left in one random variable when the other is known. The mutual information $I$ can be then interpreted as the amount of information (or reduction in uncertainty) that knowing one variable provides about the other. For example, $I(X; Y)$ is zero if $X$ and $Y$ are independent, while it is maximum when $X$ is a deterministic function of $Y$ or vice-versa.

Mutual information can be an effective measure in our case to evaluate if the input signal ($x$) can be a good predictor of the output health label ($y$). However, since computation of MI relies on estimation of probability density functions of the random variables, it is non-trivial to estimate MI robustly for high dimensional data such as the time series PPG data.
In order to overcome this curse of dimensionality, we recommend the following dimensionality reduction approaches before computing MI.

\noindent\textbf{Auto-Encoder}. Since MI is invariant under smooth invertible transformations of the variables, we propose using an auto-encoder to aggressively reduce the input space dimensionality. We train an auto-encoder with the least number of bottleneck features needed to achieve a target mean-squared reconstruction loss of $0.1$ on the normalized dataset. For the MIMIC and Aurora-BP dataset, we achieved this target with a bottleneck size of 20, at which the MI estimation worked robustly. 

\noindent\textbf{Hand-Crafted Features}. As an alternate solution to using auto-encoder, we can use hand-crafted features extracted from the input signal based on prior literature~\cite{takazawa1993clinical,elgendi2019use} and use these features for MI estimation. For example, in the task of BP prediction from PPG signal, common features include normalized systolic slope, heart rate, heart rate variability, etc. The MI estimation process helps us understand the importance of each of these features both collectively and independently. Note that in the case of hand-crafted features, there is always the concern of completeness (i.e., if the features extracted enough information from the input needed for the task), thus we recommend the auto-encoder approach whenever possible.



%% file: content/6_results.tex
\section{Analysis Details}

\subsection{Datasets}\label{dataset}

Our work builds on two datasets, the properties of which are critical to understand the results of our work.

\textbf{MIMIC II}: The MIMIC II dataset contains records of continuous high-resolution physiological waveforms of the patients in the ICU, such as ABP, PPG, and ECG sampled at 125Hz. The dataset consists of 67,830 records of varying duration from 30,000 patients~\cite{mimic_original}. For the purpose of our study, we perform our analysis on a pre-processed subset of the MIMIC II dataset, consisting of 12,000 records from 942 patients\cite{mimic}. This subset is particularly useful for our analysis as it includes a sufficient number of patients for training and testing, compliant with AAMI standards, and has been commonly utilized in previous research(Table-\ref{tab:lit_table}).

\textbf{Aurora-BP}: The Aurora-BP dataset~\cite{mieloszyk2022comparison} consists of 24,650 records from 483 subjects. Each subject has multiple records of varying duration, which was collected at rest or while performing activities like exercise and brisk walk. The records are collected from multiple sensors/devices including optical PPG, EKG, tonometer, accelerometer, and cuff-based Blood Pressure.

\subsection{Performance Standards}\label{metric}
To contextualize the performance of SBP (Systolic BP) prediction task, two benchmarks have been widely used: AAMI and BHS standards. The criteria of the AAMI (Association for the Advancement of Medical Instrumentation) standards \cite{AAMI} is that the test set should comprise of at least 85 subjects, with at least 10\% of them having an SBP above 180 mmHg and at least 10\% having an SBP under 100 mmHg.
For a test device to be compliant with the AAMI standards, the SBP prediction must have a bias under 5 mmHg and error standard deviation (SD) under 8 mmHg on the test set.
The BHS (British Hypertension Society)~\cite{bhs} standards criteria states that the test set should consist of at least 85 subjects and the cohort should be representative of the target audience of the device. The performance of the test device is divided into grades (Table-\ref{bhs-grade-table}). Additionally, the test data should cover the overall pressure range, specifically in these three ranges:
$\le$ 130, 130-160, $\ge$ 160 mmHg.

\begin{table}[H]
\begin{center}
\caption{Grading scale of test devices as per British Hypertension Society (BHS).
}
\begin{tabular}{ r c c c } 
  \toprule[1.5pt]
  & \multicolumn{3}{c}{Cumulative \% of data below SBP error threshold} \\
  Grade & $\le$ 5 mmHg & $\le$ 10 mmHg & $\le$ 15 mmHg\\
 \midrule
  A & 60\% & 85\% & 95\%  \\
  B & 50\% & 75\% & 90\% \\ 
  C & 40\% & 65\% & 85\% \\
 \bottomrule[1.5pt]
\end{tabular}
\label{bhs-grade-table}
\end{center}
\end{table}

\subsection{Other Considerations}
\begin{figure}[H]
    \includegraphics[width=\textwidth]{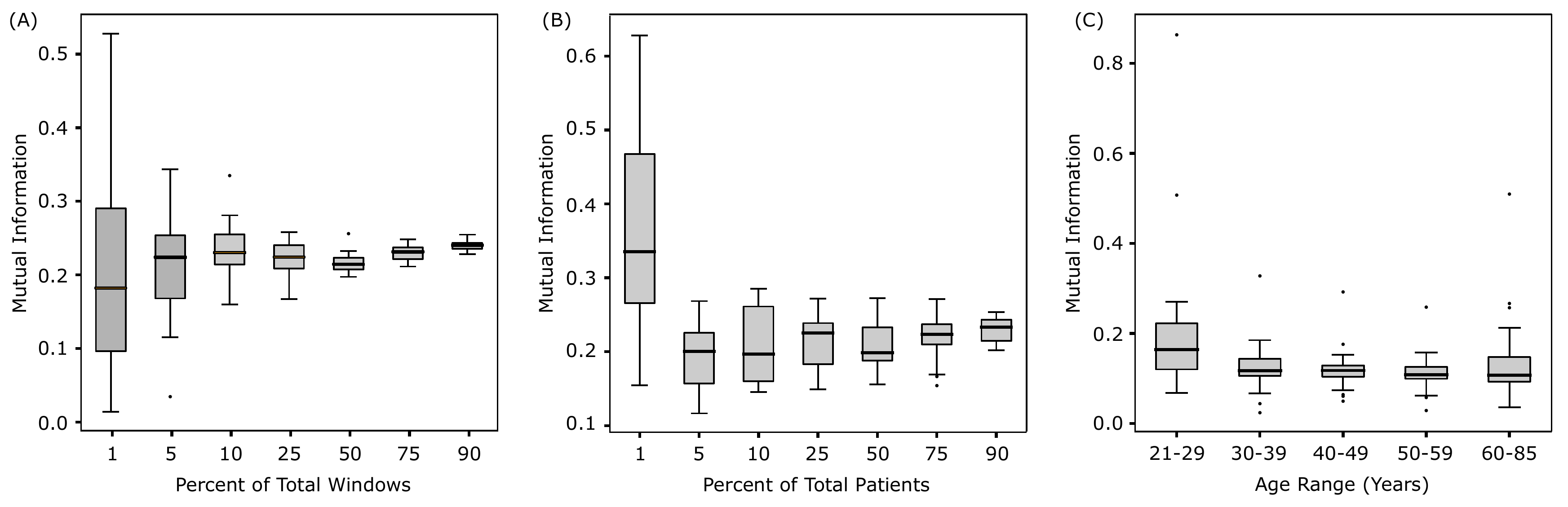}
    \caption{The effect of the number of (A) total windows (MIMIC dataset), (B) total patients (Aurora-BP dataset), as well as (C) age range (Aurora-BP dataset) on the mutual information between PPG PWA features and BP. We perform 20 runs with different random subsets of the data to plot the distributions. Optical PPG features similar to Table-\ref{tab:optical_features_results} were used for (A) and (B), while richer features (patient demographic data, PPG optical features and features derived using ECG, similar to Table-\ref{tab:feature_results}) were used for (C). For each plot the corresponding features were combined and treated as a single multi-dimensional input for computing MI.} 
    \label{fig:mi_plots}
\end{figure}

%
\textbf{Dataset size}: 
To understand the effect of data size on MI, and verify if our dataset had enough samples to enable robust MI estimation, we conducted the following experiment. We took a randomly selected slice of the data (ranging from 0.1\% to 100\% data) and computed combined MI over 20 runs (this technique is known as bootstrapping). We performed this for both the MIMIC and Aurora-BP datasets. As shown in Figures~\ref{fig:mi_plots}(A) and~\ref{fig:mi_plots}(B), although the estimates at smaller dataset sizes resulted in high variation, the variation bounds are very tight at higher sizes. This imparts confidence that our MI estimates over the full datasets are robust. Interestingly, we also found that using a smaller dataset can result in higher estimates of the MI values. This may be explained by the fact that less multi-valued mappings might be observed in a smaller sample.
Thus, having a small dataset might lead to an over optimistic perception of the relationship between input and output. 

\noindent\textbf{Participant's Demography}:
Apart from data size, we found even demographic factors, like age, impacted mutual information. Figure~\ref{fig:mi_plots}(C) shows the variation in combined MI with respect to age for the Aurora-BP dataset. In particular, we found that in the age group of 21-29 and 60-85 years, heart rate and weight were the most important features, which was not the case with the other age groups.